\documentclass[pre,twocolumn]{revtex4}

\usepackage{amsmath}
\usepackage{amssymb}
\usepackage{graphicx}

\begin{document}
\narrowtext

\title{Long-range dependence in Interest Rates and Monetary Policy}

\author{Daniel O. Cajueiro$^\star$ and Benjamin M. Tabak$^{\star\star}$}

\affiliation{$^\star$ Universidade Cat\'{o}lica de Bras\'{i}lia --
Mestrado em Economia de Empresas. \\SGAN 916, M\'odulo B -- Asa
Norte. DF 70790-160 Brazil. \\ $^{\star\star}$ Banco Central do
Brasil\\ SBS Quadra 3, Bloco B, 9 andar. DF 70074-900}

\begin{abstract}
This paper studies the dynamics of Brazilian interest rates for
short-term maturities. The paper employs developed techniques in
the econophysics literature and tests for long-range dependence in
the term structure of these interest rates for the last decade.
Empirical results suggest that the degree of long-range dependence
has changed over time due to changes in monetary policy, specially
in the short-end of the term structure of interest rates.
Therefore, we show that it is possible to identify monetary
arrangements using these techniques from econophysics.
\end{abstract}
\maketitle
%
%
%

\section{Introduction}
\label{int}

The analysis of persistence in interest rates is a fundamental
question in macroeconomics and finance. In macroeconomics, since
monetary policy is implemented through the setting of short-term
interest, the term structure of interest rates carries information
regarding expectations of future movements in short-term interest
rates. However, very little research has been conducted to study
changes in the dynamics of persistence in interest rates.

This paper presents a contribution to the literature by studying
whether changes in monetary policy stance, namely the
implementation of an inflation targeting regime and adoption of a
floating exchange rate regime, produce a change in the persistence
of interest rates. We measure persistence in this paper employing
two developed techniques in econophysics, which are the detrended
fluctuation analysis (DFA) [Moreira {\emph et al}, 2004; Peng
{\emph et al}, 2004] and Generalized Hurst exponents (GHE)
[Barab\'{a}si and Vicsek (1991)]. We show that changes in monetary
policy produce a a substantial change in persistence of interest
rates, specially for short-term interest rates.

This paper proceed as follows. In section 2 a brief review of
literature is presented. In section 3 the approaches used to
evaluate the Hurst's exponent are presented. In section 4, the
data used in this work is described. In section 5, empirical
results are presented. Finally, section 6 concludes the paper.

\section{Brief Literature Review}
\label{sec:BRL}

Recent research has studied whether long-range dependence in asset
returns and volatility changes over time, and has provided
evidence of time-varying long-range dependence (Cajueiro and
Tabak, 2004,2005). Nonetheless, while the presence of long-range
dependence in asset returns and returns volatility seems to be an
stylized fact\footnote{For details, see Mandelbrot (1971) and
Willinger \textit{et al.} (1999).}, only few papers have provided
empirical evidence of long-range dependence in interest rates
[Backus and Zin (1993), Tsay (2000), Barkoulas and Baum (1998),
McCarthy \emph{et al.} (2004), Sun and Phillips (2004), Duan and
Jacobs (1996, 2001), Cajueiro and Tabak (2006a, 2006b)].

This paper contributes to the literature by studying changes in
long-range dependence parameters in interest rates time series. We
focus on the Brazilian economy due to recent changes that occurred
in monetary policy, with the adoption of an inflation targeting
regime and a floating exchange rate regime.

\section{Measures of Long-Range Dependence}
\label{sec:lrd}

There are several methods that may be used to take into account
the long range dependence phenomena\footnote{A survey of these
methods may be found in Taqqu \emph{et al.} (1995) and Montanary
\emph{et al.} (1999). See also Hurst (1951), Lo (1991) and
Willinger \emph{et al.} (1999).}. However, in spite of the
existence of several methods, the task of calculation the Hurst
exponent is not straightforward and the methods sometimes present
incompatible estimations of the long range dependence parameter.

In this paper we follow two different approaches. The method
introduced by Barab\'{a}si and Vicsek (1991) and used recently by
Di Matteo \emph{et al.} (2005b) to measure the degree of market
development of several financial markets. According to Di Matteo
\emph{et al.} (2005b), it combines sensitivity to any type of
dependence in the data and simplicity. Moreover, since it does not
deal with $max$ and $min$ functions, it is less sensitive to
outliers than the popular R/S statistics. And, also, the detrended
fluctuation analysis (DFA) which was developed independently in
(Moreira {\emph et al}, 2004) and (Peng {\emph et al}, 2004) and
provides an alternative for the determination of the Hurst
exponent.

\subsection{Generalized Hurst exponent}

Let $Y(t)$ be the integrated time series of logarithm returns,
i.e., $Y(t)=\log{(X(t))}$. The generalized Hurst exponent is a
generalization of the approach proposed by Hurst. Barab\'{a}si and
Vicsek (1991) suggest analyzing the \textit{q}-order moments of
the distribution of increments, which seems to be a good
characterization of the statistical evolution of a stochastic
variable $Y(t)$,

\begin{equation}K_{q}(\tau)=\frac{\langle|Y(t+\tau)-Y(t)|^{q}\rangle}{\langle|Y(t)|^{q}\rangle},\end{equation}

where the time-interval $\tau$ can vary\footnote{For $q=2$, the
$K_{q}(\tau)$ is proportional to the autocorrelation function
$\rho(\tau)=\langle Y(t+\tau)Y(t)\rangle$.}. The generalized Hurst
exponent can be defined from the scaling behavior of
$K_{q}(\tau)$, which can be assumed to follow the relation

\begin{equation}K_{q}(\tau)\sim(\frac{\tau}{\nu})^{qH(q)}.\end{equation}

\subsection{Detrended fluctuation analysis}

Let $Y(t)$ be the integrated time series of logarithm returns,
i.e., $Y(t)=\log{(X(t))}$. So, one considers the
$\tau$-neighborhood around each point Y(t) of the time series. The
local trend in each $\tau$-size box is approximated by a
polynomial\footnote{This polinomial of order $m$ is usually a
first order polinomial, i.e., a straight line where the parameters
are determined by a least square fitting.} of order $m$, namely
$Z(t))$.

Then, one evaluates the local roughness, namely

\begin{equation}w^2(Y,\tau)=\frac{1}{\tau}\sum_{t\in \tau}\left(Y(t)-Z(t) \right)^2 \end{equation}

Moreira {\it et al.} (1994) showed that

\begin{equation}\langle w^2(\tau)\rangle \sim \tau^{2H} \end{equation}

\section{Data}
\label{sec:data}

The data considered here are interest rate swaps maturing on 1, 3,
6 and 12 months' time, which are the maturities available for a
``long time" span (more than 10 years of data). In these
contracts, a party pays a fixed rate over an agreed principal and
receives a floating rate over the same principal, the reverse
occurring with his or her counterpart. There are no intermediate
cash-flows, with the contracts being settled on maturity. The
floating rate is the overnight CDI rate (interbank deposits),
which tracks very closely the average rate in the market for
overnight reserves at the central bank. The fixed rate, negotiated
by the parties, is the one used on this paper.

These contracts have been traded over-the-counter in Brazil since
the early 90's, and have to be registered either on Bolsa de
Mercadorias e de Futuros - BMF (a futures exchange) or on Central
de T\'{i}tulos Privados - CETIP (a custodian).

We use data on interest rates swaps due to the lack of good
quality data on government bond indices for different maturities.
Nonetheless, these interest rates are used as benchmarks in the
Brazilian financial market.

The data is sampled daily, beginning on January 2, 1995 and ending
on May 30, 2006. The full sample has 2828 observations, collected
from the Bloomberg system.

\section{Long-Range Dependence in the Term Structure of Brazilian Interest Rates}
\label{sec:data}

This section presents results for testing for long-range
dependence in interest rates for different maturities. Table 1
shows Hurst exponents estimated using both DFA and GHE methods. We
estimate these Hurst exponents using the entire sample and also
constructing two sub-samples. The differences between monetary
policy regimes are striking, with Hurst exponents above 0.5 for
all maturities in the period before the implementation of the
inflation targeting regime and below 0.5 for maturities up to 6
months after the implementation of this regime. Panel D presents
the differences in Hurst exponents, which decrease monotonically
with maturity, suggesting that very little changed in the dynamics
of the 1 year maturity interest rate. However, changes in 1-month
interest rates were substantial, and robust to the methodology
employed to estimate Hurst exponents.

Figure 1 presents Hurst exponents for the entire sample, using
both the DFA and GHE methodology. These Hurst exponents are
decreasing but the difference in Hurst exponents is small. However
when we compare Hurst exponents across monetary regimes the
pattern changes dramatically. Figures 2 and 3 present Hurst
exponents for two different monetary regimes, using the DFA and
GHE methods.

\begin{figure}[t]
\begin{center}
\includegraphics[width=8cm,height=5cm]{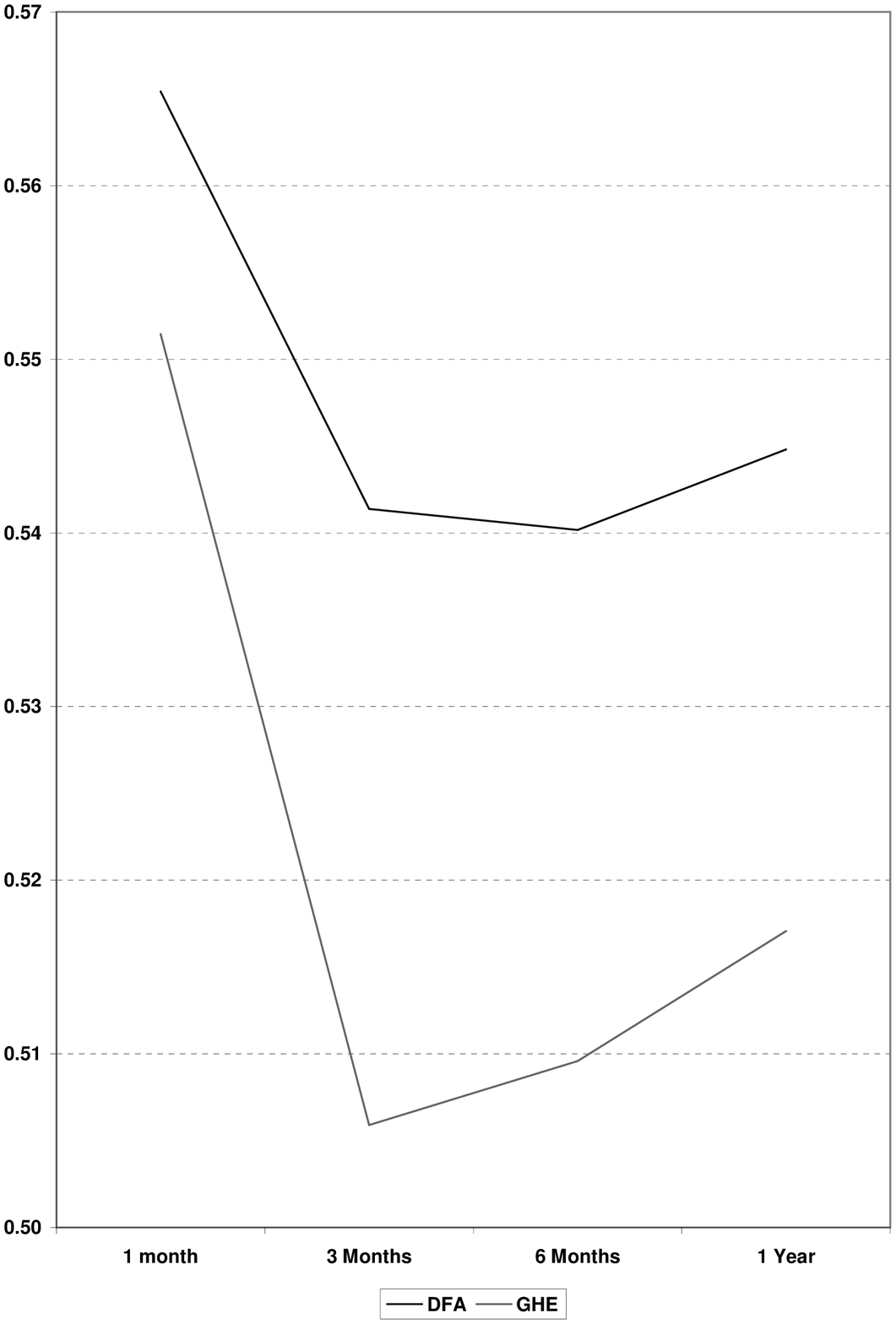}
\end{center}
\caption{Hurst exponents (DFA and GHE) for different maturity
interest rates.}\label{fig1}

\end{figure}

\begin{figure}[t]
\begin{center}
\includegraphics[width=8cm,height=5cm]{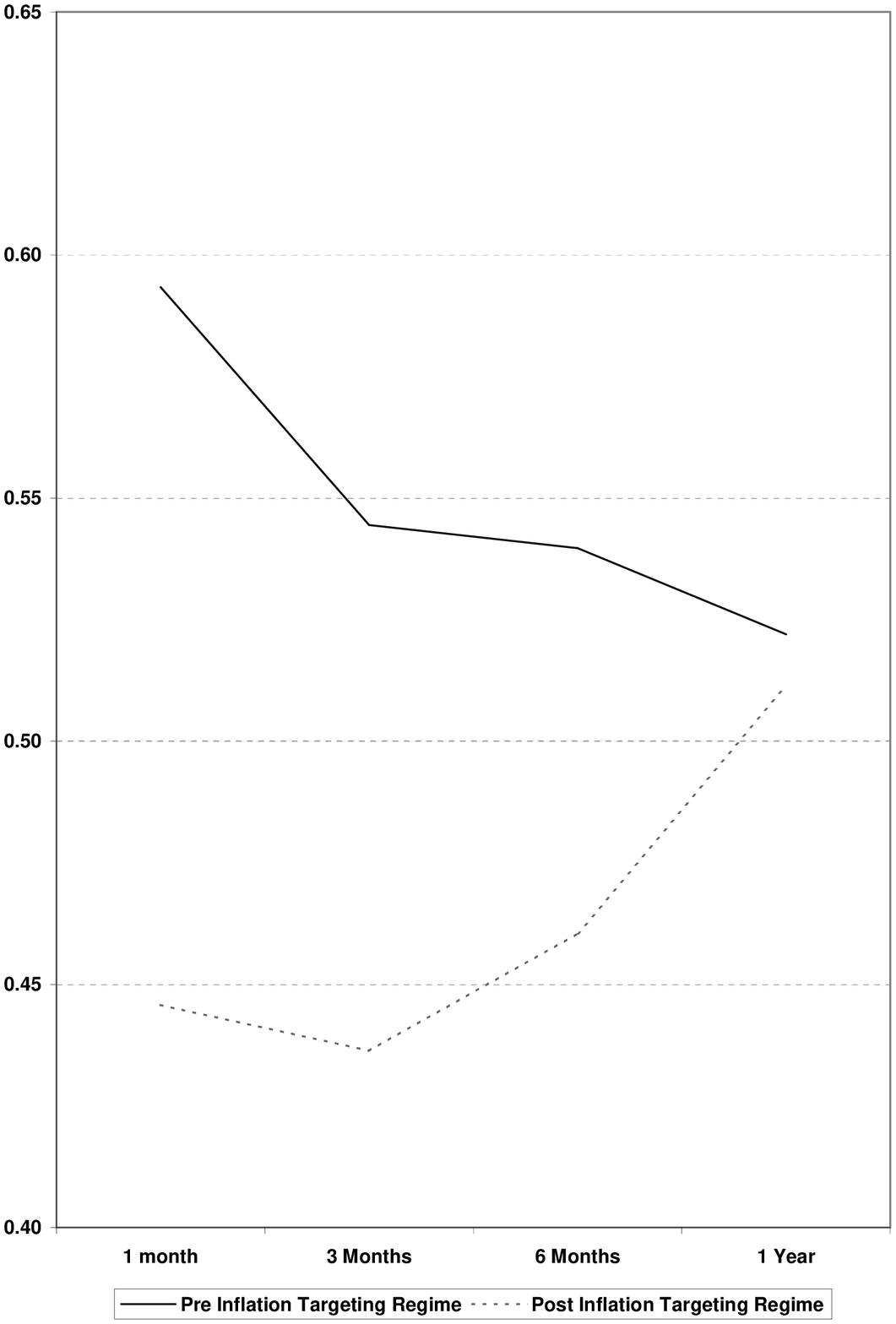}
\end{center}
\caption{Hurst exponents for different time periods
(DFA).}\label{fig2}

\end{figure}

\begin{figure}[t]
\begin{center}
\includegraphics[width=8cm,height=5cm]{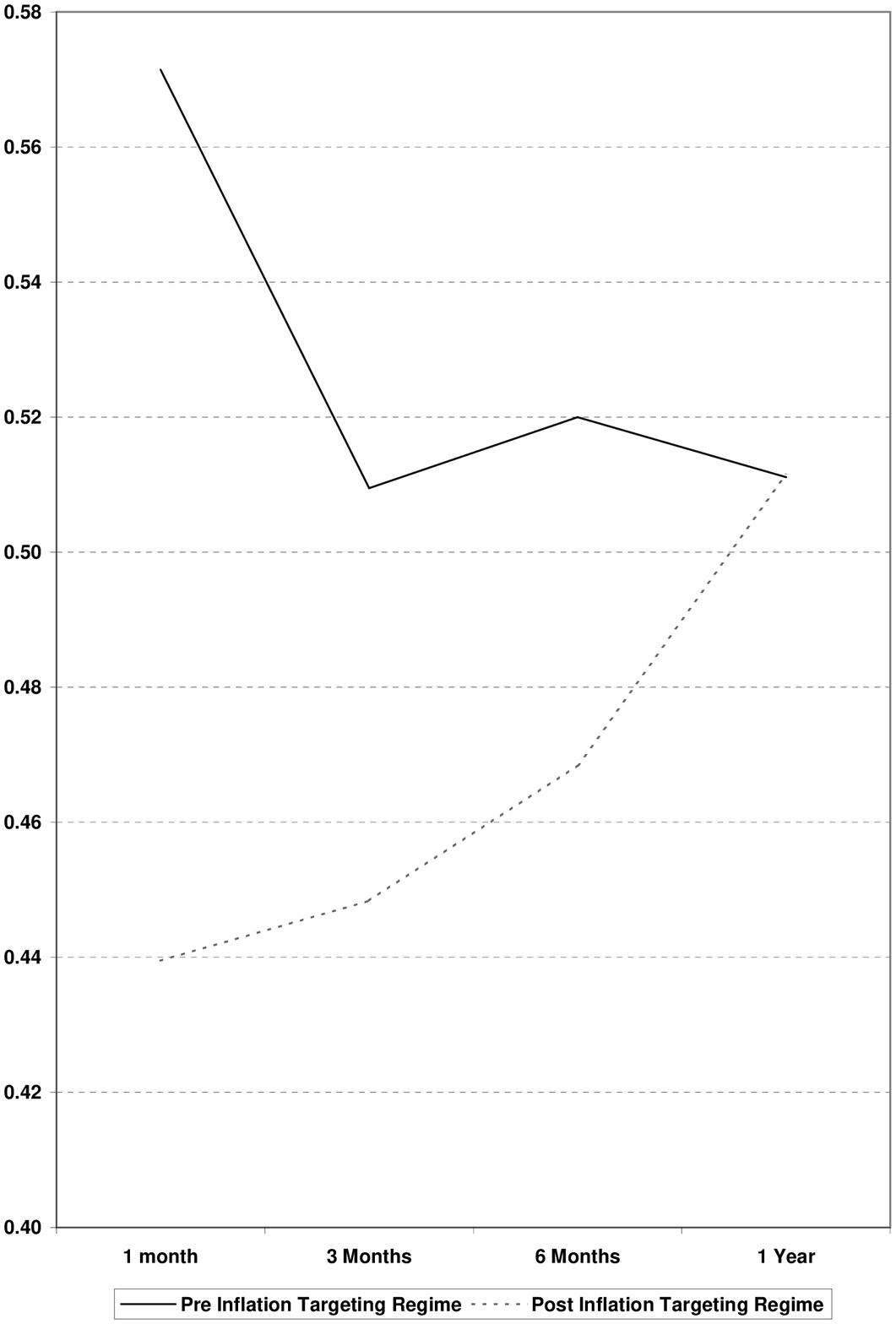}
\end{center}
\caption{Hurst exponents for different time periods
(GHE).}\label{fig3}

\end{figure}

Two main conclusions emerge from these empirical results. Hurst
exponents show a substantial break in the dynamics of persistence
in Brazilian interest rates, specially for short-term maturities.
Second, the period before the implementation of the Inflation
Targeting Regime is characterized by a downward slope in the Term
Structure of Hurst exponents, while the period after the
implementation with an upward slope. 1 year maturity interest
rates do not present any evidence of a structural break, which
suggests that studying the term structure of interest rates is
worthwhile.

\section{Conclusions}
\label{sec:conc}

In this paper we have shown empirical evidence of long-range
dependence in the Brazilian term structure of interest rates. We
show that Hurst exponents change substantially with the
implementation of the Inflation Targeting Regime in 1999, reducing
substantially interest rates persistence afterwards.

The economic intuition for the empirical results is that the
previous monetary policy cycles that occurred within the period
before the implementation of the Inflation Targeting regime have
had a substantial change. This is true because within an Inflation
Targeting regime the exchange rate is floating and therefore, the
exchange rate may absorb, at least partially, external shocks.
When exchange rates are fixed external shocks must be absorbed by
changes in domestic interest rates, and therefore, one should
expect interest rates to show more persistence.

This phenomenon is particularly true in our study, because
structural changes in interest rates were more pronounced in
short-term interest rates, with very little changes in 1 year
maturity interest rates (considered long-term interest rates in
Brazil).

Our results show that methods derived from econophysics may be
able to help explain dynamics of important macroeconomic and
financial variables such as interest rates. Further research could
focus on a variety of countries that have adopted different
monetary regimes and studying changes in persistence across
regimes.

\section{Acknowledgements} The authors thank participants of the AFPA5 2006 for helpful suggestions. Benjamin M. Tabak gratefully acknowledges financial support from CNPQ foundation. The opinions expressed in this paper are those of the authors and do not necessarily reflect those of the Banco Central do Brasil.

\clearpage
\begin{table}
\begin{tabular}{ccccc}
\hline
  &   1 month &   3 Months    &   6 Months    &   1 Year  \\

\hline
Panel A: Full Sample                                    \\
  DFA &    0.565  &    0.541  &    0.540  &    0.545  \\
GHE &    0.551  &    0.506  &    0.510  &    0.517  \\
Panel B: Pre Inflation Targeting Regime                                 \\
DFA &    0.593  &    0.544  &    0.540  &    0.522  \\
GHE &    0.571  &    0.509  &    0.520  &    0.511  \\
Panel C: Post Inflation Targeting Regime                                    \\
DFA &    0.446  &    0.436  &    0.460  &    0.511  \\
GHE &    0.440  &    0.448  &    0.468  &    0.511  \\
Panel D: Difference in Hurst Exponents ($H_{Pre}-H_{Post}$)\\
DFA &    0.148  &    0.108  &    0.079  &    0.011  \\
GHE &    0.132  &    0.061  &    0.052  &   0.000   \\

\hline
\end{tabular}
\caption{This Table presents Hurst exponents using DFA and GHE.
Results are presented for the entire sample and the pre and post
implementation of the Inflation Targeting regime.}
\label{tab:mean}
\end{table}

\end{document}